\documentclass[12pt,preprintnumbers,aps,amssymb,nofootinbib,amsmath]{revtex4}
\usepackage{epsfig,epsf}
\usepackage{graphicx}
\usepackage{epstopdf}
\newcommand{\beq}{\begin{equation}}
\newcommand{\eeq}{\end{equation}}
\def\eq#1{{(\ref{#1})}}
\def\fig#1{{Fig.~\ref{#1}}}

\newcommand{\Tr}{{\rm Tr\,}}

\newcommand{\im}{\mathrm{Im}}

\newcommand{\e}{\varepsilon}

\newcommand{\ben}{\begin{eqnarray*}}
\newcommand{\een}{\end{eqnarray*}}

\begin{document} 

\preprint{BNL-NT-07/23}
\preprint{RBRC-681}

\title{Bulk viscosity of QCD matter near the critical temperature}

\author{Dmitri Kharzeev$^a$ and Kirill Tuchin$^{b,c}$}
 
\affiliation{
a) Department of Physics, Brookhaven National Laboratory,\\
Upton, New York 11973-5000, USA\\
b) Department of Physics and Astronomy,\\
Iowa State University, Ames, Iowa, 50011, USA\\
c) RIKEN BNL Research Center,\\
Upton, New York 11973-5000, USA\\
}

\date{\today}

\begin{abstract} 
Kubo's formula relates bulk viscosity to the retarded Green's function of the trace of the energy-momentum tensor. Using low energy theorems of QCD for the latter we derive the formula which 
relates the bulk viscosity to the energy density and pressure of hot matter. We then employ 
the available lattice QCD data to extract the bulk viscosity as a function of temperature. 
We find that close to the deconfinement temperature bulk viscosity becomes large, with 
viscosity-to-entropy ratio $\zeta/s \sim 1$. 

\end{abstract}

\maketitle


One of the most striking results coming from RHIC heavy ion program is 
the observation that hot QCD matter created in $Au-Au$ collisions behaves like an almost ideal liquid rather than a gas of quarks and gluons \cite{Arsene:2004fa,Adcox:2004mh,Back:2004je,Adams:2005dq,Gyulassy:2004zy}. Indeed, hydrodynamical simulations 
of nuclear collisions at RHIC (see e.g. \cite{Teaney:2000cw,Huovinen:2001cy}) indicate that the shear viscosity of QCD plasma is very low  
even though a quantitative determination is significantly affected by the initial conditions \cite{Hirano:2005xf}. 
 This observation does not yet have any theoretical explanation due to an enormous complexity of QCD in the regime of strong coupling. This is why the information inferred from the studies of gauge theories treatable at strong coupling such as $N=4$ SUSY Yang-Mills theory is both timely and valuable. The study of shear viscosity in this theory using the holographic AdS/CFT correspondence has indicated that the shear viscosity $\eta$ at strong coupling is small, with 
 the viscosity--to--entropy ratio not far from the conjectured bound of $\eta/s = 1/4\pi$ \cite{Policastro:2001yc,Kovtun:2004de} . 
 
 \medskip
 
However  $N=4$ SUSY Yang-Mills theory is quite different from QCD; in particular it possesses 
exact conformal invariance whereas the breaking of conformal invariance in QCD is responsible for the salient features of hadronic world including the asymptotic freedom \cite{Gross:1973id}, confinement, and deconfinement phase transition at high temperature\footnote{The effects of conformal symmetry breaking on bulk viscosity of SUSY Yang-Mills theory have been studied in the framework of the AdS/CFT correspondence in ref \cite{Benincasa:2005iv}.}. 
Mathematically, conformal invariance implies the conservation of dilatational current $s_{\mu}$: 
$\partial^{\mu} s_{\mu} = 0$. Since the divergence of dilatational current in field theory is equal to the trace of the energy-momentum tensor $\partial^{\mu} s_{\mu} = \theta^{\mu}_{\mu}$, in conformally invariant theories $\theta^{\mu}_{\mu} =0$. In QCD, in the chiral limit of massless quarks the trace of the energy-momentum tensor is also equal to zero at the classical level. However quantum effects break conformal invariance \cite{scale1,scale2}:
\beq\label{trace}
\partial^\mu s_\mu=\theta^\mu_{\;\mu}=\sum_qm_q\bar qq+\frac{\beta(g)}{2g^3}
\Tr G^{\mu\nu}G_{\mu\nu}\,,
\eeq
where $\beta(g)$ is the QCD $\beta$-function, which governs the behavior of the 
running coupling: 
\beq
\mu \frac{d g(\mu)}{d \mu} = \beta (g); \label{rg}
\eeq   
note that  we have included the coupling $g$ in the definition of the gluon fields and have not written down explicitly the anomalous dimension correction to the quark mass term.

\medskip

How would this breaking of conformal invariance manifest itself in the transport properties 
of QCD plasma? How big are the effects arising from it?  The transport coefficient of the plasma which is 
directly related to its conformal properties is the bulk viscosity; indeed, it is 
 related by Kubo's formula to the correlation function of the trace of the energy-momentum tensor:
\beq\label{kubo}
\zeta = \frac{1}{9}\lim_{\omega\to 0}\frac{1}{\omega}\int_0^\infty dt \int d^3r\,e^{i\omega t}\,\langle [\theta^\mu_\mu(x),\theta^\mu_\mu(0)]\rangle \,.
\eeq
It is clear from (\ref{kubo}) that for any conformally invariant theory with $\theta^\mu_\mu \equiv \theta = 0$ the bulk viscosity should vanish. 

  The perturbative evaluation of the bulk viscosity $\zeta$ of QCD plasma has been performed recently \cite{Arnold:2006fz}, and yielded a very small value, with  $\zeta / s \sim 10^{-3}$. The parametric smallness of bulk viscosity can be easily understood from eqs (\ref{kubo}) and (\ref{trace}) which show that $\zeta \sim \alpha_s^2$, in accord with the result of ref. \cite{Arnold:2006fz}. This would seem to suggest that bulk viscosity effects in the quark-gluon plasma are unimportant. 
However, perturbative expansions at temperatures close to the critical one are not applicable, so at moderate temperatures 
one has to rely on lattice QCD calculations. Lattice calculations of the equation of state become increasingly precise; however, the direct calculations of transport coefficients have been notoriously difficult.  
Two calculations have been reported for shear viscosity \cite{Nakamura:2004sy,Meyer:2007ic}, including a recent high statistics study \cite{Meyer:2007ic}. Both indicate that $\eta/s$ is not much higher than the conjectured bound of $1/4\pi$; no lattice calculations of the bulk viscosity have been reported so far.
Fortunately, the correlation function of the trace of the energy-momentum tensor in QCD is 
constrained by the low-energy theorems, which do not rely on perturbation theory. They can thus be used to express the bulk viscosity in terms of the ``interaction measure" $\left< \theta \right> = \mathcal{E}-3P$ where $\mathcal{E}$ is the energy density and $P$ is the pressure,  which are 
measured on the lattice with high precision. Such a study is the subject of this Letter.
 
 \medskip

The calculation of the bulk viscosity starts with the Kubo's formula (\ref{kubo}) (we follow the definitions and notations of  \cite{LL9}).
Introducing the retarded Green's function we can re-write (\ref{kubo}) as
\beq\label{kubo2}
\zeta=\frac{1}{9}\lim_{\omega\to 0}\frac{1}{\omega}\int_0^\infty dt\int d^3r\,e^{i\omega t} \, iG^R(x)\,=\,\frac{1}{9}\lim_{\omega\to 0}\frac{1}{\omega} \,iG^R(\omega,\vec 0)\,=\,-\frac{1}{9}\lim_{\omega\to 0}\frac{1}{\omega} \,\im G^R(\omega,\vec 0)\,.
\eeq
The last equation follows from the fact that due to P-invariance, function $\im G^R(\omega,\vec 0)$ is odd in $\omega$ while $\mathrm{Re}\,G^R(\omega,\vec 0)$ is even in $\omega$.  Let us define the spectral density
\beq\label{spec.den}
\rho(\omega,\vec p)=-\frac{1}{\pi}\,\im G^R(\omega,\vec p)\,.
\eeq
Using the Kramers-Kronig relation the retarded Green's function can be represented 
as
\beq\label{kk}
G^R(\omega,\vec p)=\frac{1}{\pi}\int_{-\infty}^\infty\,\frac{\im G^R(u,\vec p)}{u-\omega -i\e}\,du=
\int_{-\infty}^\infty\,\frac{\rho(u,\vec p)}{\omega -u+i\e}\,du
\eeq

The retarded  Green's function $G^R(\omega,\vec p)$ of a bosonic excitation is related to the Euclidean Green's function $G^E(\omega,\vec p)$ by analytic continuation
\beq\label{analit}
G^E(\omega,\vec p)=-G^R(i\omega,\vec p)\,,\quad \omega >0\,.
\eeq
Using \eq{kk} and the fact that $\rho(\omega,\vec p)=-\rho(-\omega,\vec p)$ we recover
\beq\label{ge00}
G^E(0,\vec 0)=2\int_0^\infty \frac{\rho(u,\vec 0)}{u}\,du\,.
\eeq

\medskip

As we discussed above, the scale symmetry of QCD lagrangian is broken by quantum vacuum fluctuations. As a result  the trace of the energy momentum tensor $\theta$ acquires a non-zero vacuum expectation value. The correlation functions constructed out of operators $\theta(x)$ satisfy a chain of low energy theorems (LET) which are a consequence of the renormalization group invariance of observable quantities \cite{Novikov:1981xj}. These low-energy theorems entirely determine the dynamics of the effective low-energy theory. This effective theory has an elegant geometrical interpretation \cite{Migdal:1982jp}; in particular, gluodynamics can be represented as a classical theory formulated on a curved (conformally flat) space-time background \cite{KLT-Pom}.   At finite temperature, the breaking of scale invariance by quantum fluctuations results in $\theta=\mathcal{E}-3P\neq 0$ clearly observed on the lattice for $SU(3)$ gluodynamics \cite{Boyd:1996bx}; the presence of quarks \cite{Bernard:1996cs} including the physical case of two light and a strange quark \cite{Karsch:2007vw}, or  
considering large $N_c$ \cite{Bringoltz:2005rr} does not change this conclusion. 

\medskip

The LET of  Ref.~\cite{Novikov:1981xj,Migdal:1982jp} were generalized to the case of finite temperature in \cite{Ellis:1998kj,Shushpanov:1998ce}. The lowest in the chain of relations reads (at zero baryon chemical potential):
\beq\label{let}
G^E(0,\vec 0)=\int d^4x\,\langle T\theta(x),\theta(0)\rangle =\left(T\frac{\partial}{\partial T} -4\right)\,\langle \theta\rangle_T\,.
\eeq
To relate the thermal expectation value of $\langle \theta\rangle_T$ to the quantity $(\mathcal{E}-3P)_\mathrm{LAT}$ computed on the lattice, we should keep in mind that
\beq\label{renor}
(\mathcal{E}-3P)_\mathrm{LAT} = \langle \theta\rangle_T - \langle \theta\rangle_{0},
\eeq
i.e. the zero-temperature expectation value of the trace of the energy-momentum tensor
\beq
\langle \theta\rangle_{0} = - 4 |\epsilon_v|
\eeq
has to be subtracted; it is related to the vacuum energy density $\epsilon_v < 0$.
Now, using \eq{ge00}, \eq{let} and \eq{renor} we derive the following sum rule
\beq\label{sum}
2\int_0^{\infty} \frac{\rho(u,\vec 0)}{u}\,du=-\left(4-T\frac{\partial}{\partial T} \right)\,\langle \theta\rangle_T=T^5\frac{\partial}{\partial T}\frac{(\mathcal{E}-3P)_\mathrm{LAT}}{T^4}+16|\epsilon_v|\,,
\eeq
 This {\it exact} relation is the main result of our paper. 

\medskip
\begin{figure}[ht] 
      \includegraphics[width=10cm]{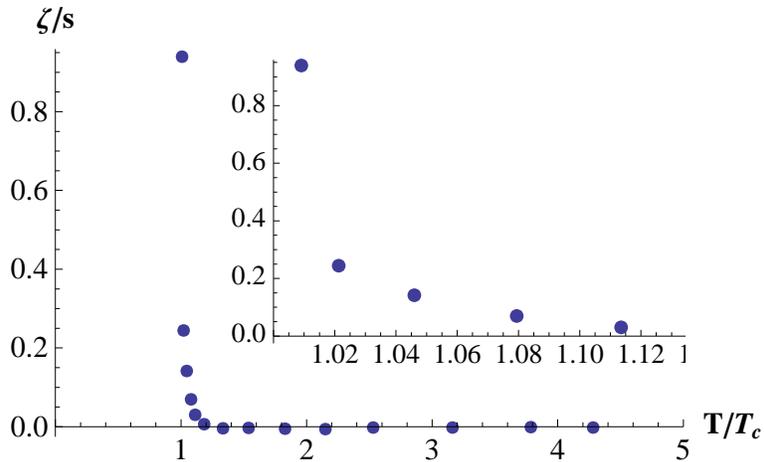} 
\caption{The ratio of bulk viscosity to the entropy density for $SU(3)$ gluodynamics. We have used $|\epsilon_v|=0.62\, T_c^4$ and  $T_c=0.28$ GeV \cite{Boyd:1996bx}.}
\label{fig:z}
\end{figure}

In order to extract the bulk viscosity $\zeta$ from (\ref{sum}) we need to make an ansatz for the spectral 
density $\rho$. At high frequency, the spectral density should be described by perturbation theory; however, 
the perturbative (divergent) contribution has been subtracted in the definition of the quantities on the r.h.s.\ 
of the sum rule (\ref{sum}), and so we should not include the perturbative continuum\footnote{For an explicit perturbative expression and a discussion of the properties of $\rho(u)$ at small frequencies see e.g. \cite{Fujii:1999xn}.} $\rho(u) \sim \alpha_s^2\, u^4$ on the l.h.s.\ as well.   
In the small frequency region, we will {\it assume}  
 the following ansatz
\beq\label{ansatz}
\frac{\rho(\omega,\vec 0)}{\omega}=\frac{9\,\zeta}{\pi}\frac{\omega_0^2}{\omega_0^2+\omega^2}\,,
\eeq
which satisfies \eq{kubo2} and \eq{spec.den}. Substituting (\ref{ansatz}) in \eq{sum} we arrive at 
\beq\label{ze}
\,\zeta= \frac{1}{9\,\omega_0}\left\{ T^5\frac{\partial}{\partial T}\frac{(\mathcal{E}-3P)_\mathrm{LAT}}{T^4}+16|\epsilon_v|\right\}\,.
\eeq
A peculiar feature of this result is that the bulk viscosity is linear in the difference $\mathcal{E}-3P$, rather than 
quadratic as naively implied by the Kubo's formula. This is similar to the strong coupling result obtained 
for the non-conformal supersymmetric Yang-Mills gauge plasma \cite{Benincasa:2005iv}.

\medskip

The parameter $\omega_0 = \omega_0(T)$ is a scale at which the perturbation theory becomes valid. On dimensional grounds, we expect it to be proportional to the temperature, $\omega_0  \sim T$. We estimate it as the scale at which the lattice calculations of the running coupling \cite{Kaczmarek:2004gv} coincide with the perturbative expression at a given temperature. In the region $1< T/T_c <3$ we find $\omega\approx  (T/T_c) \ 1.4\ {\rm GeV}$. Now we are ready to use (\ref{ze}) to extract the bulk viscosity from the lattice data.  

The results of the numerical calculation using as an input the high precision lattice data \cite{Boyd:1996bx} are displayed in \fig{fig:z}. One can see that away from $T_c$ the bulk viscosity is small, in accord with the expectations based on the perturbative results \cite{Arnold:2006fz}. However, close to $T_c$ the rapid growth 
of $\mathcal{E}-3P$ causes a dramatic increase of bulk viscosity. Basing on the lattice results \cite{Nakamura:2004sy,Meyer:2007ic} which indicate that the shear viscosity remains small close to $T_c$, we expect that bulk viscosity will be the dominant correction to the ideal hydrodynamical behavior in the vicinity of the deconfinement phase transition.  
 
\medskip

We thank F. Karsch for providing us with the numerical lattice data and helpful discussions. We acknowledge useful conversations with A. Buchel, J. Ellis, J. Kapusta, R. Lacey, G. Moore, A. Starinets, D. Teaney and A. Vainshtein. The work of D.K. was supported by the U.S. Department of Energy under Contract No. DE-AC02-98CH10886. K.T. is grateful to RIKEN, BNL, and the U.S. Department of Energy (Contract No. DE-AC02-98CH10886) for providing facilities essential for the completion of this work.


\end{document}